\documentclass[12pt]{article}
\usepackage{amsmath}
\usepackage{graphicx}
\usepackage{natbib}
\usepackage{url} 
\usepackage{psfrag}
\usepackage{natbib}

\usepackage{array}
\usepackage{mathrsfs} 
\usepackage{amsfonts, amssymb, latexsym, slashbox}
\usepackage{amsmath,amsthm}
\usepackage{cancel}
\usepackage{graphics, bm}
\usepackage{hyperref}
\usepackage{changes}  
\usepackage{comment}
\def\em{\it}
\def\emph{\textit}
\input{colordvi}

\def\o*{o_{{\!}_{P^*}}}
\def\O*{\cal O_{{\!}_{P^*}}}

\def\E{\mathrm{E}}

\def\P{\mathrm{Pr}}

\def\B{\mathrm{B}}
\def\P{\mathrm{P}}

\def\.{\mbox{.}}

\def\sfrac(#1,#2){\mbox{$\frac{#1}{#2}$}}

\def\T{\mathrm{\scriptscriptstyle\top}}

\def\ceiling#1{\left\lceil #1 \right\rceil}
\def\ints(#1,#2){\mathbb{I}_{#1}^{#2}}
\newtheorem {theorem}{Theorem}
\newtheorem {lemma}{Lemma}
\newtheorem {proposition}{Proposition}
\newtheorem {rem}{Remark}
\newtheorem{assump}{}

\newcommand{\blind}{0}

\begin{document}

\def\spacingset#1{\renewcommand{\baselinestretch}%
{#1}\small\normalsize} \spacingset{1}


\if0\blind
{
  \title{\bf Maximum Approximate Bernstein Likelihood Estimation of Densities in a Two-sample Semiparametric Model}
  \author{Zhong Guan
\\
    Department of Mathematical Sciences\\
Indiana University South Bend}
  \maketitle
} \fi

\if1\blind
{
  \bigskip
  \bigskip
  \bigskip
  \begin{center}
    {\LARGE\bf Maximum Approximate Bernstein Likelihood Estimation of Densities in a Two-sample Semiparametric Model}
\end{center}
  \medskip
} \fi

\bigskip
\begin{abstract}
Maximum likelihood estimators  are proposed for the parameters and the densities in a semiparametric density ratio model in which the nonparametric baseline density is  approximated by the Bernstein polynomial model.  The EM algorithm  is used to obtain the maximum approximate Bernstein likelihood estimates. Simulation study shows that the performance of the proposed method is much better than the existing ones. The proposed method is illustrated by real data examples. Some asymptotic results are also presented and proved.
\end{abstract}

\noindent%
\vfill

\newpage
\spacingset{1.5} 
\section{Introduction}
\label{sect: introduction}
Nonparametric density estimation is a difficult task in statistics. It is even more difficult for small sample data. For each $x$ in the support of a density $f$ in a nonparametric
 model, the information for this one-dimensional parameter $f(x)$ is zero (see \cite{Bickel-etal-1993-book}).
\cite{Ibragimov-Hasminskii-1982} also showed that there is no nonparametric model
for which this information is positive. Properly reducing the infinite dimensional parameter to a finite dimensional one is necessary.  To estimate an unknown smooth function as the nonparametric component of a non- and semi-parametric model, as we have done in empirical likelihood we usually approximate it by a step-function and parameterize it using the jump sizes of the step-function.
This approach gives an efficient estimate of the underlying cumulative distribution function. Because this estimate is a step-function, we have to use kernel or other method to smooth it to obtain a density estimate. However, kernel density is actually the {\em convolution} of the scaled kernel and the underlying distribution to be estimated. There is always trade-off between the bias and variance. In semiparametric problems, the roughness of the step-function approximation could also affect the finite sample performance of the estimates of the parametric components.  Instead of approximating the underlying distribution function by a step-function and then smoothing the discretized estimation, \cite{Guan-jns-2015} proposed to use
a Bernstein polynomial approximation and to directly and smoothly  estimate the underlying distribution using a maximum approximate Bernstein likelihood method. \cite{Guan-jns-2015}'s method parameterizes the underlying distribution by the coefficients of the Bernstein polynomial and differs from other Bernstein polynomial smoothing methods which was initiated by \cite{Vitale1975} and use empirical distribution to estimate these coefficients. The maximum approximate Bernstein likelihood method has been successfully applied to grouped, contaminated, multivariate, and interval censored data \citep{Guan-2017-jns,Guan-2019-mable-deconvolution, Wang-and-Guan-2019, Guan-SIM-2020}. In application to the Cox's proportional hazards regression model, not only a smooth estimate of the survival function but also improved estimates of regression coefficients can be resulted, due to a better approximation of the unknown underlying baseline density function.

In applications of statistics especially in biostatistics, independent two-sample data from case-control study for instance  are common. If the two nonparametric underlying
distributions are linked in a certain parametric way, then we can find better estimates of the distributions by efficiently combining the two independent samples. Examples of such linked models are two-sample proportional odds model \citep{Dabrowska-Doksum-JASA-1988}, two-sample proportional hazard model \citep{Cox1972}, two-sample density ratio (DR) model
\citep[see for example,][]{Qin-Zhang-1997-bka,Qin-and-Zhang-2005,CHENG-and-CHU-Bernoulli-2004}, and so on.
 Suppose that the densities $f_0$ and $f_1$ of ``control'' data $X_0$ and ``case'' data $X_1$, respectively, satisfy the following density ratio model
\begin{equation}\label{eq: exponential tilting model}
f_1(x)=f(x;\bm\alpha)=f_0(x)\exp\{\bm\alpha^\T \tilde r(x)\},
\end{equation}
where $\bm\alpha=(\alpha_0,\ldots,\alpha_d)^\T \in\mathcal{A}\subset R^{d+1}$, and $\tilde r(x)=(1, r^\T(x))^\T$. In this model $f_0$ is also called ``baseline'' density.
%
  Let $D$ be a binary
response variable, $\pi_j=\P(D=j)$,
$j=0,1$.  Define  $f_i(x)=f_{X|D}(x|D=i)$, $j=0,1$. By Bayes' theorem, the two-sample DR model
is equivalent to the following  {\em logistic regression model}
\citep{Qin-Zhang-1997-bka}
\begin{equation}\label{eq: Logistic Reg}
\log\left\{\frac{P(D=1|X=x)}{P(D=0|X=x)}\right\}= {\bm\alpha}^{*\T}\tilde r(x),
\end{equation}
where $ \alpha_0^*=\alpha_0-\log(\pi_0/\pi_1)$ and  {$\alpha_i^*=\alpha_i$}, $i\ge 1$. Model (\ref{eq: exponential tilting model}) is appropriate because the right-hand side of (\ref{eq: Logistic Reg}) can be a good approximation of the log odds function.  The goodness-of-fit of this model is also testable \citep{Qin-Zhang-1997-bka}. An advantage of this model is that one can also choose $f_1$ as the baseline density, that is, $f_0(x)= f_1(x)\exp\{-\bm\alpha^\T \tilde r(x)\}$. For transformed data $Y=h(X)$ we have $g_1(y)=g_0(y)\exp\{\bm\alpha^\T \tilde r[h^{-1}(y)]\}$, where 
$h^{-1}(\cdot)$ is the inverse of $h(\cdot)$ and $g_i(y)$ is the density of $Y$ given $D=i$, $i=0,1$.
Model (\ref{eq: exponential tilting model}) was also used for one-sample density estimation by \cite{Efron-and-Tibshirani-1996-bim} in which $f_0$ is a carrier density and $r(x)$ is a  known $d$-dimensional sufficient statistic.

Parametrizing the infinite dimensional parameter $f_0$ in (\ref{eq: exponential tilting model}) using the multinomial model with unknown probabilities at the observations results in the maximum empirical likelihood estimator (MELE)  \citep{Qin-Zhang-1997-bka} $\tilde{\bm\alpha}$ of $\bm\alpha$ and step-function estimator of $f_0$.
 The MELE  $\tilde{\bm\alpha}$ can also be obtained by fitting the data $(X,D)$ with the logistic regression (\ref{eq: Logistic Reg}). This method works well when $f_0$ is a nuisance parameter. However in many applications, both $\bm\alpha$ and $f_0$ are of interest. A jagged step-function estimate of $f_0$ is unsatisfactory especially when sample sizes are small. Smooth and efficient estimator is desirable.
\cite{Qin-and-Zhang-2005} proposed to smooth the discrete empirical density estimates of $f_0$ and $f_1$ using  {kernel method}. As a smoothing technique kernel density does not target the unknown density but its convolution with the scaled kernel function for any positive bandwidth.  Good density estimation is key to solve many difficult statistical problems such as the goodness-of-fit test \citep{CHENG-and-CHU-Bernoulli-2004} and the estimation of the receiver
operating characteristic curve when result of diagnostic test is continuous \citep{Zou-etal-1997-sim} and sample size is small.
 In this paper, we shall investigate the estimation of densities and the parameters under model (\ref{eq: exponential tilting model}) using approximate Bernstein likelihood method.

The nonparametric component $f_0$ in the semiparametric model is  totally unspecified.  If we have no information about the support
 of $f_0$, we can only estimate $f_0$ as a density with support $[z_{(1)}, z_{(n)}]$, where $z_{(1)}$ and $z_{(n)}$ are, respectively, the minimum and maximum order statistics of a pooled sample of size $n$ from $f_0$ and $f_1$.
 If the density $f_i$  of $X_i$ has support $[a,b]$, $i=0,1$, and $f_1(x)=f_0(x)\exp\{\bm\alpha^{\T} \tilde r(x)\}$,  then the desnity of $Y_i=(X_i-a)/(b-a)$ is $g_i(y)=(b-a)f_i[a+(b-a)y]$ which have support $[0,1]$ and satisfy
 $g_1(y)
 =g_0(y)\exp\{\bm\alpha^{\T}\tilde r[a+(b-a)y]\}$.
Without loss of generality we will assume that both $f_0$ and $f_1$ have support $[0,1]$.

The paper is organized as follows. The approximate Bernstein polynomial model for DR model is introduced and is proved to be nested in Section \ref{sect: Methodology}.
The EM algorithm for finding the maximum approximate Bernstein likelihood estimates of the mixture proportions and the regression coefficients, the methods for determining a lower bound for the model degree $m$ based on sample mean and variance and for choosing the optimal degree $m$ are also given in this section.
The proposed methods are illustrated by some real datasets in Section \ref{example} and  compared with some
existing competitors through Monte Carlo experiments in Section \ref{sect: simulations}.  Some asymptotic results
about the convergence rate of the proposed estimators are presented in Section \ref{sect: large sample property}.  Some concluding remarks are given in Section \ref{set: concluding remarks}. The proofs of the theoretical results are relegated  to the Appendix.
\section{Methodology}
\label{sect: Methodology}
\subsection{Approximate Bernstein Polynomial Model}
Let $\bm x_{n_i}=\{x_{i1},\ldots,x_{in_i}\}$ be independent observations of $X_i$, $i=0,1$. The true loglikelihood is
$\ell(\bm\alpha,f_0)=\ell(\bm\alpha,f_0; \bm z_n)=\sum_{i=1}^n\log f_0(z_i)+\bm\alpha^\T \sum_{j=1}^{n_1}\tilde  r(x_{1j}),$ where $\bm z_n=\{z_1,\ldots,z_n\}=\{x_{01},\ldots,x_{0n_0}$; $x_{11},\ldots,x_{1n_1}\}$, $n=n_0+n_1$.
Define simplex $\mathbb{S}_m=\{(u_0,\ldots,u_m): u_i\ge 0, \sum_{i=0}^m u_i=1\}$.
Instead of discretizing baseline density $f_0$ with finite
support $\bm z_n$ as in \cite{Qin-Zhang-1997-bka}, we use Bernstein polynomial approximation
\citep{Guan-jns-2015}
$f_0(x)\approx f_m(x;\bm 0,\bm p)= f_m(x;\bm p)= \sum_{j=0}^m p_j \beta_{mj}(x),$
where $\bm p\in \mathbb{S}_m$,  
and $\beta_{mj}(x)=(m+1){m\choose j}x^j(1-x)^{m-j}$ is the density of beta distribution with shape parameters $(j+1,m-j+1)$, $j\in \mathbb{I}_0^m$. Here  and in what follows $\ints(m,n)=\{m,\ldots,n\}$ for any integers $m\le n$.
Therefore
$f(x;\bm\alpha)$ can be approximated by $f_m(x;\bm\alpha,\bm p)\equiv f_m(x;\bm p)\exp\{\bm\alpha^\T \tilde r(x)\}$.
The cumulative distribution function of $f_m(x;\bm\alpha,\bm p)$ is
$F_m(x; \bm\alpha,\bm p)=\sum_{j=0}^m p_j B_{mj}(x;\bm\alpha)$, where
$B_{mj}(x;\bm\alpha)=\int_0^x \beta_{mj}(y)\exp\{\bm\alpha^\T \tilde r(y)\}dy.$
The approximate loglikelihood is then
\begin{equation}\label{eq: bernstein loglik for raw data}
\ell_m(\bm\alpha,\bm p)=\sum_{i=1}^{n}\log f_m(z_i;\bm p)+\bm\alpha^\T \sum_{j=1}^{n_1}\tilde  r(x_{1j}),
\end{equation}
with 
constraint
\begin{equation}\label{eq: constraints}
(\bm\alpha,\bm p)\in\Theta_m(\mathcal{A})\equiv \biggr\{(\bm\alpha,\bm p)\in\mathcal{A}\times \mathbb{S}_m:  \sum_{i=0}^mp_{i}w_{mi}(\bm\alpha)=1\biggr\},
\end{equation}
where
$w_{mj}(\bm\alpha)= B_{mj}(1;\bm\alpha)$, 
$j\in\ints(0,m).$
Under constraint (\ref{eq: constraints}) the approximate density $f_m(x;\bm\alpha,\bm p)$ is mixture of $\beta_{mj}(x; \bm\alpha)=\beta_{mj}(x)\exp\{\bm\alpha^\T \tilde r(x)\}/w_{mj}(\bm\alpha)$  with mixing proportions
$\tilde p_j(\bm\alpha)=p_jw_{mj}(\bm\alpha)$,
$j\in\ints(0,m)$.

For the given $r(\cdot)$ and $\mathcal{A}$, let $\mathcal{D}_m(\mathcal{A})$ be the family of all functions
$f_m(x;\bm\alpha,\bm p_m)= f_m(x;\bm p_m)\exp\{\bm\alpha^\T \tilde r(x)\}$, 
 $(\bm\alpha,\bm p_m)\in \Theta_m(\mathcal{A})$. The following proposition implies that the models $\mathcal{D}_m(\mathcal{A})$ are nested.
\begin{proposition}\label{prop: two-sample BPDR models are nested}
For the given regressor vector $r(\cdot)$ and parameter space $\mathcal{A}$,  $\mathcal{D}_m(\mathcal{A})\subset\mathcal{D}_{m+1}(\mathcal{A})$, for all positive integers $m$.
\end{proposition}

The maximizer $\hat{\bm\theta}=(\hat{\bm\alpha},\hat{\bm p})$ of $\ell_m(\bm\alpha,\bm p)$  subject to constraint (\ref{eq: constraints}) for an optimal degree $m$ is called the {\em maximum approximate Bernstein likelihood estimate} (MABLE) of $\bm\theta=(\bm\alpha, \bm p)$.
Then $f_i$ and $F_i$, respectively,  can be estimated
by $\hat f_{i}(x)= f_m(x;i\hat{\bm\alpha},\hat{\bm p})$ and $\hat F_{i}(x)=F_m(x;i\hat{\bm\alpha},\hat{\bm p})$, $i=0,1$.

For densities $f_i$ on $[a,b]$ which satisfy (\ref{eq: exponential tilting model}), we can obtain $(\hat{\bm\alpha}, \hat{\bm p})$ based on transformed data $(x_{ij}-a)/(b-a)$  with $r(x)$ replaced by $r[a+(b-a)x]$. Then we have estimates of $f_i$ and $F_i$, respectively,
\begin{align}\label{eq: MABLE of fi, i=0,1}
\hat f_{i}(x)=&~\frac{1}{b-a}f_m\Big(\frac{x-a}{b-a}; i\hat{\bm\alpha},\hat{\bm p}\Big)=\frac{\exp\{i\hat{\bm\alpha}^\T\tilde r(x)\}}{b-a} \sum_{j=0}^m \hat p_j \beta_{mj}\Big(\frac{x-a}{b-a}\Big),\\
 \label{eq: MABLE of Fi, i=0,1}
\hat F_{i}(x)=&~F_m\Big(\frac{x-a}{b-a}; i\hat{\bm \alpha},\hat{\bm p}\Big)=\sum_{j=0}^m \hat p_j B_{mj}\Big(\frac{x-a}{b-a};i\hat{\bm \alpha}\Big),~\mbox{$i=0,1$,}
\end{align}
 where
$B_{mj}(x;  {\bm \alpha})=\int_0^x \beta_{mj}(u)\exp\{{\bm \alpha}^{\T} \tilde r[a+(b-a)u]\}du$, $x\in[0,1]$,  $j\in \ints(0,m)$.

To find maximum likelihood estimates of the parameters $(\bm\alpha, \bm p)$
we first introduce some notations. For any function $\varphi(\bm\alpha)$ which may also depend upon the data, its first and second derivatives with respect to $\bm\alpha$ are denoted by
$\dot\varphi(\bm\alpha)=\frac{\partial \varphi(\bm\alpha)}{\partial \bm\alpha}$ and $\ddot \varphi(\bm\alpha)=\frac{\partial^2 \varphi(\bm\alpha)}{\partial \bm\alpha\partial \bm\alpha^\T}.$
The entries are denoted by
$[\dot\varphi(\bm\alpha)]_i=\frac{\partial \varphi(\bm\alpha)}{\partial \alpha_i}$ and  $[\ddot \varphi(\bm\alpha)]_{ij}=\frac{\partial^2 \varphi(\bm\alpha)}{\partial \alpha_i\partial \alpha_j}$,  $i,j\in\ints(0,d).$
For example, the derivatives of $w_{mj}(\bm\alpha)$, $j\in\ints(0,m)$, are
$\dot w_{mj}(\bm\alpha)
= \int_0^1 \tilde r(x)\beta_{mj}(x)\exp\{\bm\alpha^\T \tilde r(x)\} dx$ and
$\ddot w_{mj}(\bm\alpha)
= \int_0^1\tilde r(x)\tilde r^\T(x)\beta_{mj}(x)\exp\{\bm\alpha^\T \tilde r(x)\} dx$.
Note $\tilde r(x)=(1, r^\T(x))^\T$, $[\dot w_{mj}(\bm\alpha)]_0=w_{mj}(\bm\alpha)$,  $[\ddot w_{mj}(\bm\alpha)]_{00}=w_{mj}(\bm\alpha)$,
and $[\ddot w_{mj}(\bm\alpha)]_{0i}=[\dot w_{mj}(\bm\alpha)]_i$, $i\in\ints(0,d)$.

The standard EM algorithm combined with method of Lagrange multipliers leads to the following algorithm.

\paragraph{Algorithm for finding $(\hat{\bm\alpha},\hat{\bm p})$ with a given $m$:}
\begin{itemize}
  \item[] 
  \begin{itemize}
  \item [Step 0.] Choose small numbers $\epsilon_1, \epsilon_2>0$ and large integers $N_1$ and $N_2$.
  \item [Step 1.] Use the logistic regression to find an MELE $\bm\alpha^{(0)}=\tilde{\bm\alpha}$. Choose a uniform initial $p^{(0)}=\bm 1^\T/(m+1)$ for $\bm p$. If vanishing boundary contraints $f_0(0)=0$ and/or $f_0(1)=0$ are available, choose $p_0^{(0)}=0$ and/or $p_m^{(0)}=0$ accordingly and set the other $p_i$'s uniformly.
  \item [Step 2.] Set $s=0$, $\bm\theta^{(s)}=(\bm\alpha^{(s)},\bm p^{(s)})$. Calculate $\ell_m^{(s)}=\ell_m(\bm\theta^{(s)})=\ell_m(\bm\alpha^{(s)},\bm p^{(s)})$.
  \item [Step 3.] Set $t=0$, 
  $\bm\alpha^{\langle t\rangle}=\tilde{\bm\alpha}$.
  Run the Newton-Raphson iteration
 $\bm\alpha^{\langle t+1 \rangle} =
    \bm\alpha^{\langle t\rangle}
-J_s^{-1}(\bm\alpha^{\langle t\rangle})
\bm H_s(\bm\alpha^{\langle t\rangle})$, $t=0,1,2,\ldots,$ until $|\bm\alpha^{\langle t+1 \rangle}-
    \bm\alpha^{\langle t\rangle}|<\epsilon_1$ or $t>N_1$ to obtain $\bm\alpha^{(s+1)}=\bm\alpha^{\langle t+1 \rangle}$, where
\begin{align}
\label{eq for alpha simplified group}
H_s(\bm\alpha)&= \sum_{j=1}^{n_1} \tilde r(x_{1j})-n_1 \sum_{k=0}^m  \frac{T_k(\bm \theta^{(s)})\dot w_{mk}(\bm\alpha)}{n_0+n_1 w_{mk}(\bm\alpha)},
\\
J_s(\bm\alpha)
\label{eq: Jacobian for raw data}
   &=    -n_1\sum_{k=0}^m
  \frac{[n_0+n_1w_{k}(\bm\alpha)]\ddot w_{k}(\bm\alpha)-n_1\dot w_{k}(\bm\alpha)\dot w_{k}^\T(\bm\alpha)}{[n_0+n_1w_{mk}(\bm\alpha)]^2} T_{k}(\bm \theta^{(s)}),\;\;\;
\\
 \label{eq: Tk for raw data}
T_k(\bm\theta^{(s)})&
=\sum_{i=0}^1\sum_{j=1}^{n_i}\frac{p_k\beta_{mk}(x_{ij})}{f_m(x_{ij}; \bm p)},\quad k\in\ints(0,m).
\end{align}

  \item [Step 4.] set
$p_k^{(s+1)}= p_k(\bm\alpha^{(s+1)},\bm \theta^{(s)})$, $k\in\ints(0,m)$,
where
\begin{equation}\label{eq: pk(alpha,p(s)) for grouped data}
p_k(\bm\alpha,\bm \theta^{(s)})
=\frac{T_k(\bm\theta^{(s)})}{n_0+n_1 w_{mk}(\bm\alpha)},\quad k\in\ints(0,m).
\end{equation}

  \item [Step 5.] Set $s=s+1$.  Calculate $\ell_m^{(s)}=\ell_m(\bm\alpha^{(s)},\bm p^{(s)})$.
  \item [Step 6.]  If $\ell_m^{(s)}-\ell_m^{(s-1)}<\epsilon_2$ or $s>N_2$ then set $\hat{\bm\theta}=(\hat{\bm\alpha}, \hat{\bm p})=(\bm\alpha^{(s)},\bm p^{(s)})$ and stop. Otherwise  go to Step 3.
\end{itemize}
\end{itemize}
Bootstrap method can be used to approximate the standard error of $\hat{\bm\alpha}$:
 Generate $x_{i1}^*,\ldots,x_{in_i}^*$ from $\hat f_{i}(x)$, $i=0,1$, and fit the bootstrap samples by the proposed model with $m=\hat m$ or $\tilde m$ to obtain $\hat{\bm\alpha}^*$.  Repeat the boostrap run a large number of times and estimate the standard error of $\hat{\bm\alpha}$ by the sample standard deviation of $\hat{\bm\alpha}^*$.

\subsection{Choice of baseline and the model degree}\label{sect: optimal degree}
Let $\hat m_b^{(i)}=\max\{\ceiling{\bar x_i(1-\bar x_i)/s^2_i-3},1\}$ be the estimated lower bound for $m$  based $x_{ij}$, $j=1,\ldots,n_i$ as in \cite{Guan-jns-2015,Guan-2017-jns}. If $\hat m_b^{(1)}<\hat m_b^{(0)}$ we switch ``case'' and ``control'' data and take $f_1$ as baseline so that the estimated lower bound for the model degree of the two-sample density ratio model is $\hat m_b=\min\{\hat m_b^{(0)}, \hat m_b^{(1)}\}$. Proposition \ref{prop: two-sample BPDR models are nested} implies that $\ell_m(\hat{\bm\alpha}, \hat {\bm p})$ is nondecreasing in $m$. Applying the change-point method of \cite{Guan-jns-2015} to $\ell_m(\hat{\bm\alpha}, \hat{\bm p})$ one can obtain an optimal degree $\hat m$.  In many cases an optimal degree is very close to $\hat m_b$. The search of an optimal degree starts at some $m_0 <\hat m_b$.
Approximating $\ell_m(\hat{\bm\alpha}, \hat{\bm p})$ by $\ell_m(\tilde{\bm\alpha}, \tilde{\bm p})=\max_{\bm p\in\Theta_m(\{\tilde{\bm\alpha}\})}\ell_m(\tilde{\bm\alpha},\bm p)$, where $\tilde{\bm\alpha}$ is the MELE  of $\bm\alpha$, can reduce the cost of EM computation and results in an optimal degree  $\tilde m$.
One can obtain
$\tilde{\bm p}$  by iteration
\begin{equation}\label{eq: iteration for p-tilde for grouped data}
\tilde p_k^{(s+1)}=\frac{T_k(\tilde{\bm\alpha},\tilde {\bm p}^{(s)})}{n+\lambda^{(s)}(\tilde{\bm\alpha})[ w_{mk}(\tilde{\bm\alpha})-1]},\quad k\in\ints(0,m),\quad s\in\ints(0,\infty),
\end{equation}
where $T_k(\bm\alpha,\bm p)$ is given by (\ref{eq: Tk for raw data})
and $\lambda=\lambda^{(s)}(\tilde{\bm\alpha})$ can be obtained by Newton-Raphson iteration 
$\lambda^{\langle t+1 \rangle}=\lambda^{\langle t \rangle}- {\psi(\lambda^{\langle t \rangle})}/{\psi'(\lambda^{\langle t \rangle})}$, $t\in\ints(0,\infty)$,
where
\begin{align*}
\psi(\lambda)&=\sum_{k=0}^m  p_k(\tilde{\bm\alpha},  \bm p^{(s)})\left[w_{mk}(\tilde{\bm\alpha})-1\right]=\sum_{k=0}^m  \frac{T_k(\tilde{\bm\alpha}, \bm p^{(s)})\left[w_{mk}(\tilde{\bm\alpha})-1\right] }{n+\lambda\left[w_{mk}(\tilde{\bm\alpha})-1\right]},\\
\psi'(\lambda)&= -\sum_{k=0}^m  \frac{T_k(\tilde{\bm\alpha}, \bm p^{(s)})\left[w_{mk}(\tilde{\bm\alpha})-1\right]^2}{\{n+\lambda\left[w_{mk}(\tilde{\bm\alpha})-1\right]\}^2}.
\end{align*}
%
The proposal is implemented in R as a component of package \texttt{mable} \citep{mable} which is publically available.

\section{Real Data Application}\label{example}
\subsection{Coronary Heart Disease Data}   
\begin{figure}
\centering
  \makebox{\includegraphics[width=5.5in]{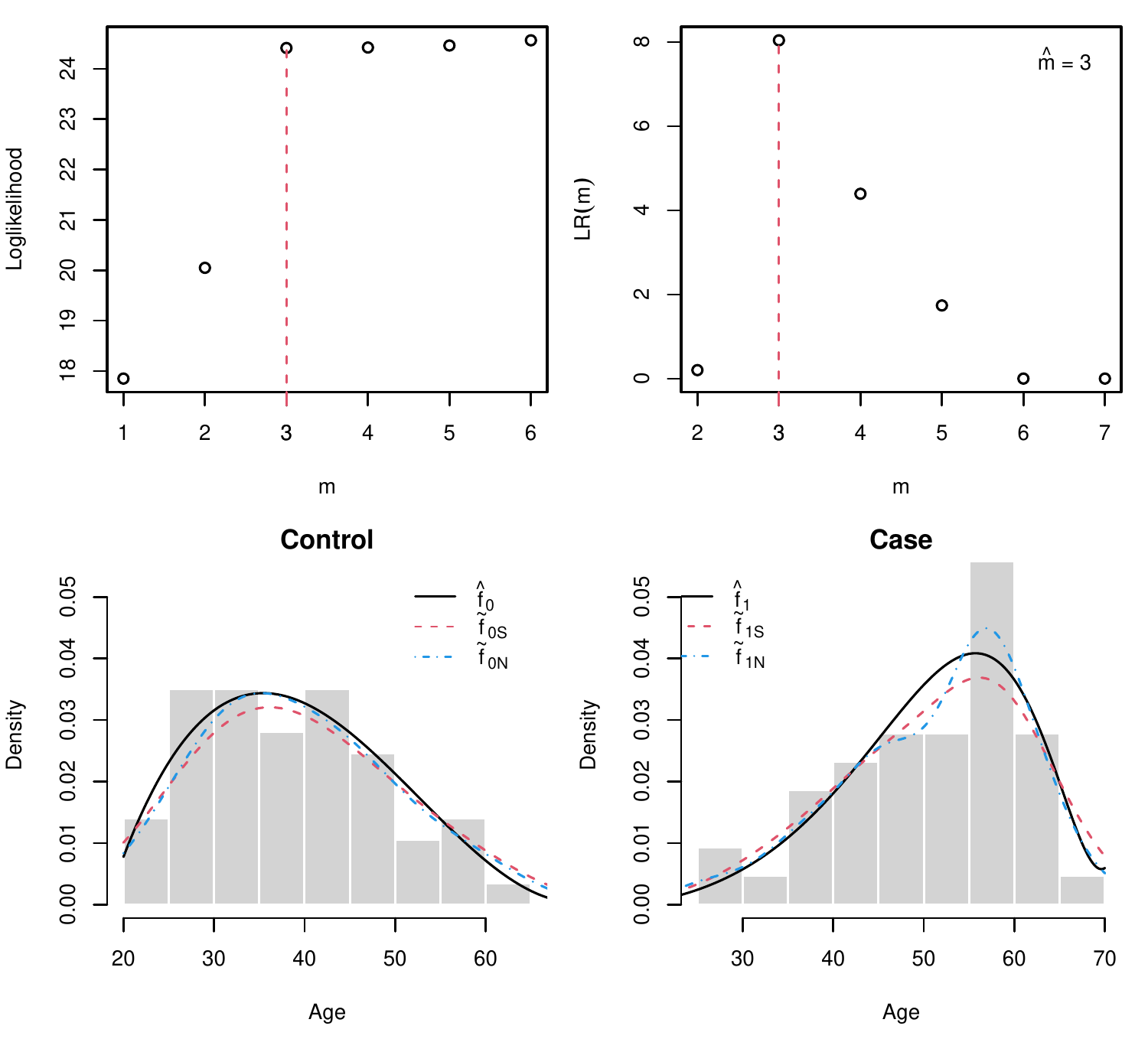}}
\caption{Coronary Heart Disease  Data. Upper left panel: log-likelihood of the data; upper right panel: likelihood ratio for change-point estimate.
Lower panel: histograms(light gray),  the MABLE $\hat f_{i}$,  the semparametric kernel density estimate $\tilde f_{i\mathrm{S}}$, and  the one-sample nonparametric kernel density estimate $\tilde f_{i\mathrm{N}}$ of $f_i$, $i=0,1$.
}\label{fig:Example1}
\end{figure}

\cite{Hosmer1989} analyzed the
relationship between age and the status of coronary heart disease (CHD)
based on 100 subjects participated in a study.  The data set contains $n_0=57$ ages from control group and  $n_1=43$  ages from case group:
$y_0=$(20, 23, 24, 25, 26, 26, 28, 28, 29, 30, 30, 30, 30, 30, 32,
32, 33, 33, 34, 34, 34, 34, 35, 35, 36, 36, 37, 37, 38, 38, 39,
40, 41, 41, 42, 42, 42, 43, 43, 44, 44, 45, 46, 47, 47, 48, 49,
49, 50, 51, 52, 55, 57, 57, 58, 60, 64)  and
$y_1=$(25, 30, 34, 36, 37, 39, 40, 42, 43, 44, 44, 45, 46, 47, 48,
48, 49, 50, 52, 53, 53, 54, 55, 55, 56, 56, 56, 57, 57, 57, 57,
58, 58, 59, 59, 60, 61, 62, 62, 63, 64, 65, 69).
The extreme sample statistics are $z_{(1)}=20$ and $z_{(n)}=69$.
We choose truncation interval $[a,b]=[20,70]$, $r(y)=y$, and transform $y_i$'s  to $x_i=(y_i-a)/(b-a)$, $i=0,1$. The control is selected as baseline and $\hat m_b=3$. Using $M=\ints(1, 20)$ as a candidate set we obtained $\tilde m=\hat m=3$ (see the upper panel of Figure  \ref{fig:Example1}).
The MABLE's of $f_i$ and $F_1$ are given by (\ref{eq: MABLE of fi, i=0,1}) and (\ref{eq: MABLE of Fi, i=0,1}) with  $\hat{\bm p}=(0.09686, 0.89834, 0.00000, 0.004796)^\T$ and $\hat{\bm\alpha}=(-5.040, 0.111)^\T$ with SE
$(0.945, 0.020)^\T$ 
based 1000 bootstrap runs. This is very close to the MELE $\tilde{\bm\alpha}=(-5.02760,0.11092)^\T$ with
SE $(1.134, 0.024)^\T$ \citep{Hosmer1989,Qin-and-Zhang-2005}.

The lower panel of Figure \ref{fig:Example1} also shows the
   proposed density estimates, the semiparametric estimates of \cite{Qin-and-Zhang-2005} based on two-sample empirical likelihood method with Gaussian
kernel and the  kernel density estimates using Gaussian kernel based on one sample only. We can see
that the proposed method gives a smoother density estimate.
 From Figure \ref{fig:Example1} we see that the MABLE's $\hat f_{i}$ differs from the other two density estimates of $f_i$ especially especially for the case data. The $\hat f_{0}$ leans a little bit more to the left.  All estimates of $f_1$  show
 strong evidence supporting the observation that individuals at age between 45 and 60 are more likely to have CHD.

\subsection{Pancreatic Cancer Data} 
We apply the proposed method to the Pancreatic cancer diagnostic marker data in which sera from $n_0 = 51$ control patients with pancreatitis and $n_1 = 90$
case patients with pancreatic cancer were studied at the Mayo
Clinic with a cancer antigen, CA-125, and with a carbohydrate
antigen, CA19-9. \cite{Wieand-et-al-1989-bka} showed that CA19-9 has higher sensitivity to Pancreatic cancer. Let $y_{ij}$, $j=1,\ldots,n_i$, $i=0,1$, denote the logarithm of the observed value of CA19-9 for the $j$th subject of control group ($i=0$) and case group ($i=1$). The combined sample is $\{z_1,\ldots,z_n\}$, $n=n_0+n_1$. \cite{Qin-Zhang-Biometrika-2003} considered the measurement $y$ on CA19-9  and obtained $p$-value 0.769 of the Kolmogorov--Smirnov--test for the
density ratio model with $r(y)=(y, y^2)^\T$. \cite{Qin-Zhang-Biometrika-2003}'s MELE is $\tilde{\bm\alpha}=(0.56, -1.91, 0.45)^\T$ with SE $(1.66, 1.22, 0.21)^\T$. 

We choose $a=z_{(1)}=0.8754687$ and $b=z_{(n)}=10.08581$. The estimated lower bounds for $m$ based on ``control'' and ``case'' data are, respectively, $\hat m_b^{(0)}=19$ and $\hat m_b^{(1)}=2$. We chose ``case'' as baseline.
Based on the transformed data $x_{ij}=(y_{ij}-a)/(b-a)$, we obtain an optimal degree
$\tilde m=\hat m = 3$ and estimates $\hat f_i$, $i=0,1$, as given by (\ref{eq: MABLE of fi, i=0,1}) with $m=3$, where
$\hat{\bm\alpha}=(0.045,  -1.677, 0.434)^\T$  with SE $(1.35, 0.91, 0.15)^\T$  based 1000 bootstrap runs,
%
%
and
$(\hat p_0,\ldots,\hat p_3)=(0.09747, 0.42829, 0.38557, 0.08867)$.
%
The case density estimates agree each other. These results show that healthy people have lower logarithmic level of CA 19-9 in their blood while logarithmic levels of CA 19-9 for pancreatic cancer patients are nearly uniform.
\begin{figure}
\centering
  \makebox{\includegraphics[width=6.5in]{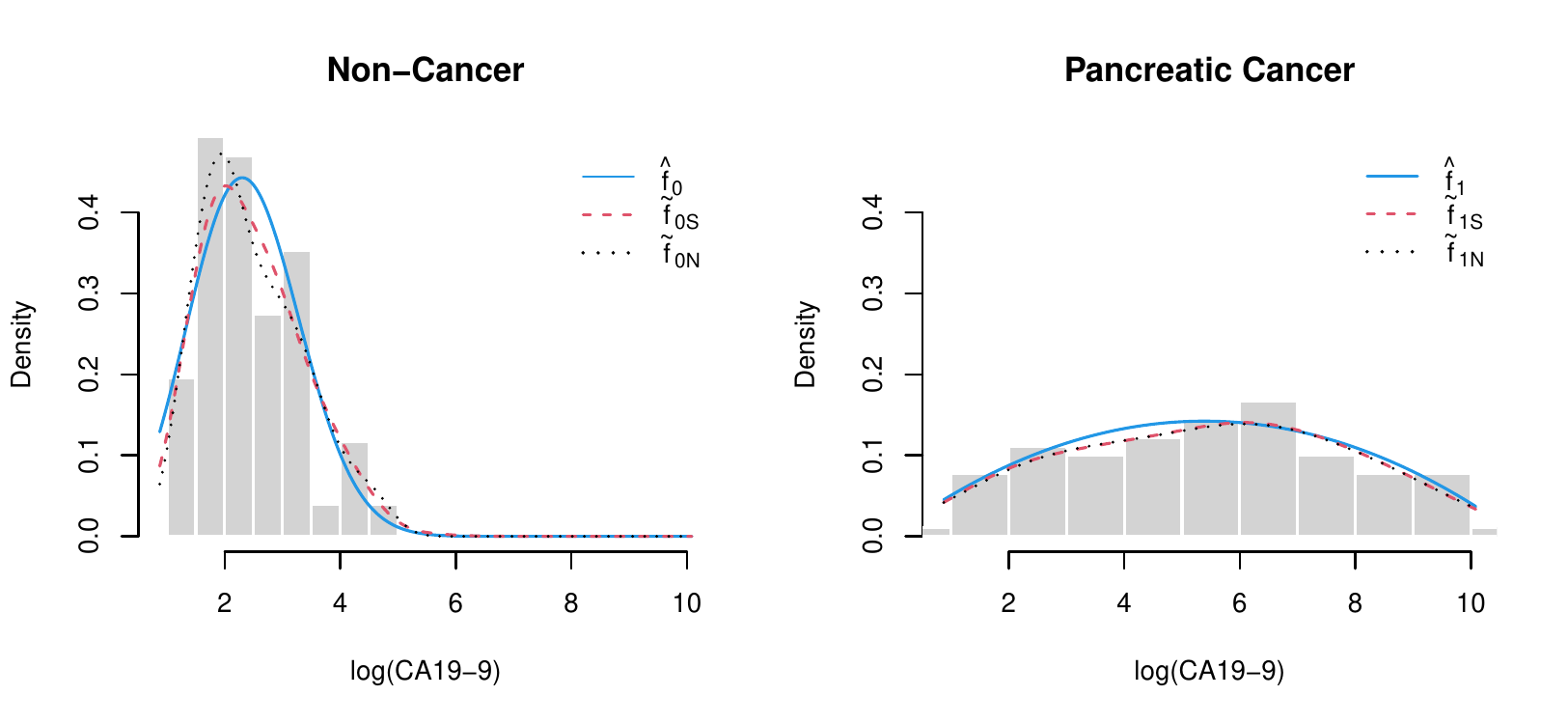}}
\caption{Pancreatic cancer CA 19-9 data. Histograms(light gray)  and density estimates of log CA 19-9 level without pancreatic cancer (left panel) and with pancreatic cancer (right panel):  the MABLE $\hat f_{i}$,  the semparametric kernel density estimate $\tilde f_{i\mathrm{S}}$, and the one-sample nonparametric kernel density estimate $\tilde f_{i\mathrm{N}}$ of $f_i$, $i=0,1$.
}\label{fig:Example2-densities}
\end{figure}

\subsection{Melanoma Data} \cite{Venkatraman-and-Begg-bka-1996} compared two systems which can be used to evaluate suspicious lesions of being a melanoma based on
 paired data.  The two systems are the clinical
score system given by doctors  and the dermoscope.
\cite{Qin-Zhang-Biometrika-2003} suggest the density ratio model with $r(x)=x$.
The MELE of $\bm\alpha$ is $\tilde{\bm\alpha}=(0.887, 1.000)$ with SE $(0.37,0.23)^\T$.

\begin{figure}
\centering
  \makebox{\includegraphics[width=5.5in]{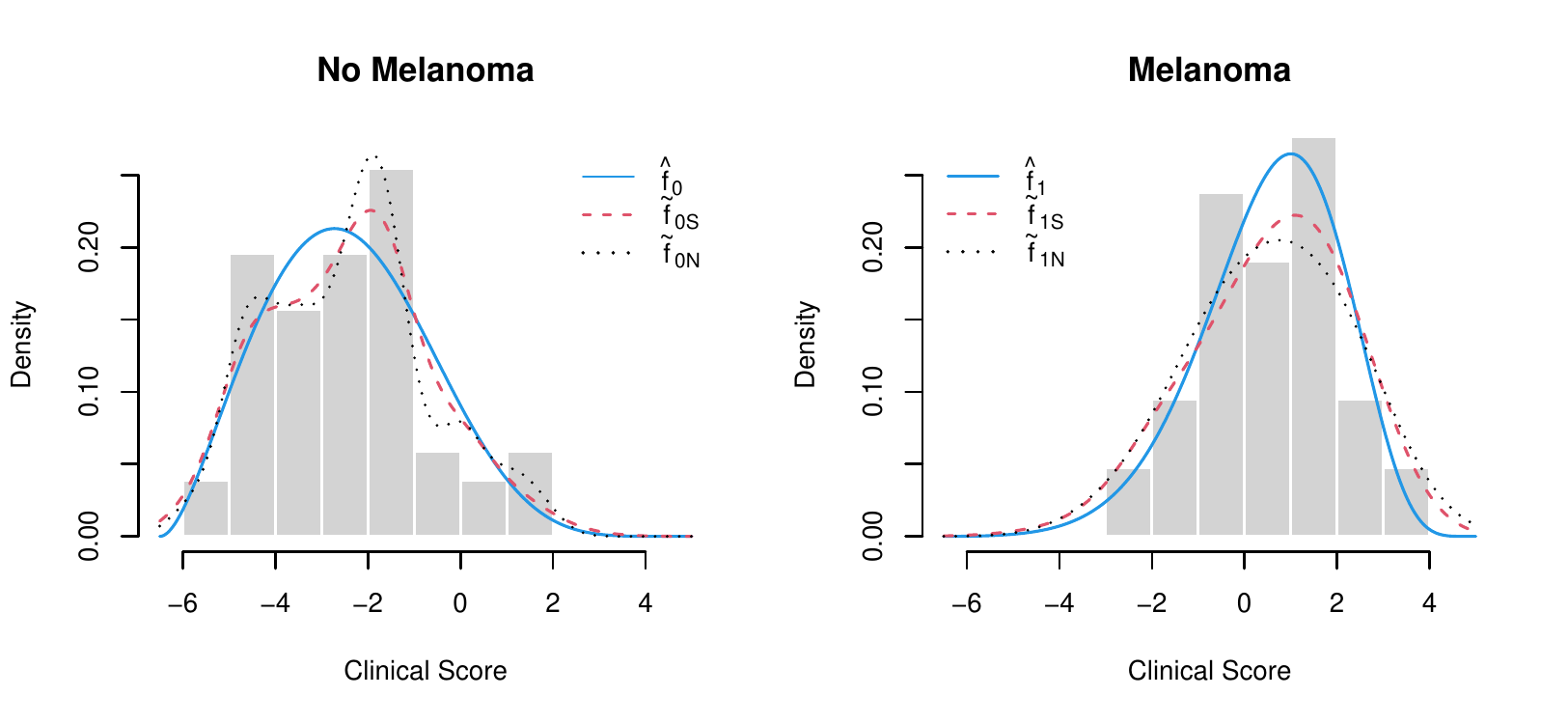}}
\caption{Melanoma data. Histograms(light gray)  and density estimates of clinical scores without melanoma (left panel) and with melanoma (right panel): the MABLE $\hat f_{i}$, the semparametric kernel density estimate $\tilde f_{i\mathrm{S}}$, and the one-sample nonparametric kernel density estimate $\tilde f_{i\mathrm{N}}$ of $f_i$, $i=0,1$.
}\label{fig:melanoma data densities}
\end{figure}

Using the proposed method with 
$a=z_{(1)}=-6.5$ and $b=5.0$ we have model degree $\hat m=\tilde m=10$.
We obtained the MABLE $\hat{\bm\alpha}= (0.881, 1.018)^\T$ of $\bm\alpha$ with SE $(0.75,0.77)^\T$ based 1000 bootstrap runs,   $\hat p_i<10^{-5}$, $i\notin \ints(2,5)$, and
 $(\hat p_2,\cdots,\hat p_5)=(.30153, .14619, .55226, .00002)$.
From Figure \ref{fig:melanoma data densities} we see that  the clinical scores have different distributions with a small overlap for people with and without melanoma.

\section{Simulation}\label{sect: simulations}
In this section we compare the performances of the proposed estimator $\hat f_{i}$ with the one-sample parametric MLE $\hat f_{i\mathrm{P}}$, the two-sample semiparametric estimator
$\tilde f_{i\mathrm{S}}$ of \cite{Qin-and-Zhang-2005}, and the one-sample kernel density estimator $\tilde f_{i\mathrm{N}}$  by examining the point-wise
mean squared error (pMSE) $\mathrm{mse}_j$ at
$t_j=a+j(b-a)/N$, $j\in\ints(0,N)$, $N=512$) and approximate mean integrated squared error (MISE) $\mathrm{mise}=N^{-1}\sum_{j=1}^N \mathrm{mse}_j$ for $i=0,1$. For convenience and fair comparison, we used same setups as in \cite{Qin-and-Zhang-2005}.
 The sample were generated using the models below. In all the simulations, the sample sizes are $(n_0, n_1) = (50,50), (100, 100)$ and the number of Monte Carlo runs is 1000.

{\bf Model 1: Normal distributions} $X_0\sim N(0,1)$, $X_1\sim N(\mu,1)$, $\tilde r(x)=(1,x)^\T$,   $f_0(x)=\frac{1}{\sqrt{2\pi}}e^{-x^2/2}$, $\bm\alpha=(-\mu^2/2, \; \mu)^\T$, and
%
 $\mu=0.25(0.25)2.00$. We choose $a=\min(-4, \mu-4)$ and $b=\max(4,\mu+4)$. In this model, the bandwidths for $\tilde f_{i\mathrm{S}}$ and $\tilde f_{i\mathrm{N}}$ are those suggested by \cite{Qin-and-Zhang-2005}. The parametric MLE  is $\hat f_{i\mathrm{P}}(x)=f_0[(x-\bar x_i)/s_i]/s_i$, where $\bar x_i$ and $s_i$ are, respectively, the sample mean and sample standard deviation of $x_{i1},\ldots,x_{in_i}$, $i=0,1$.
\begin{figure}
\centering
  \makebox{\includegraphics[width=5.8in]{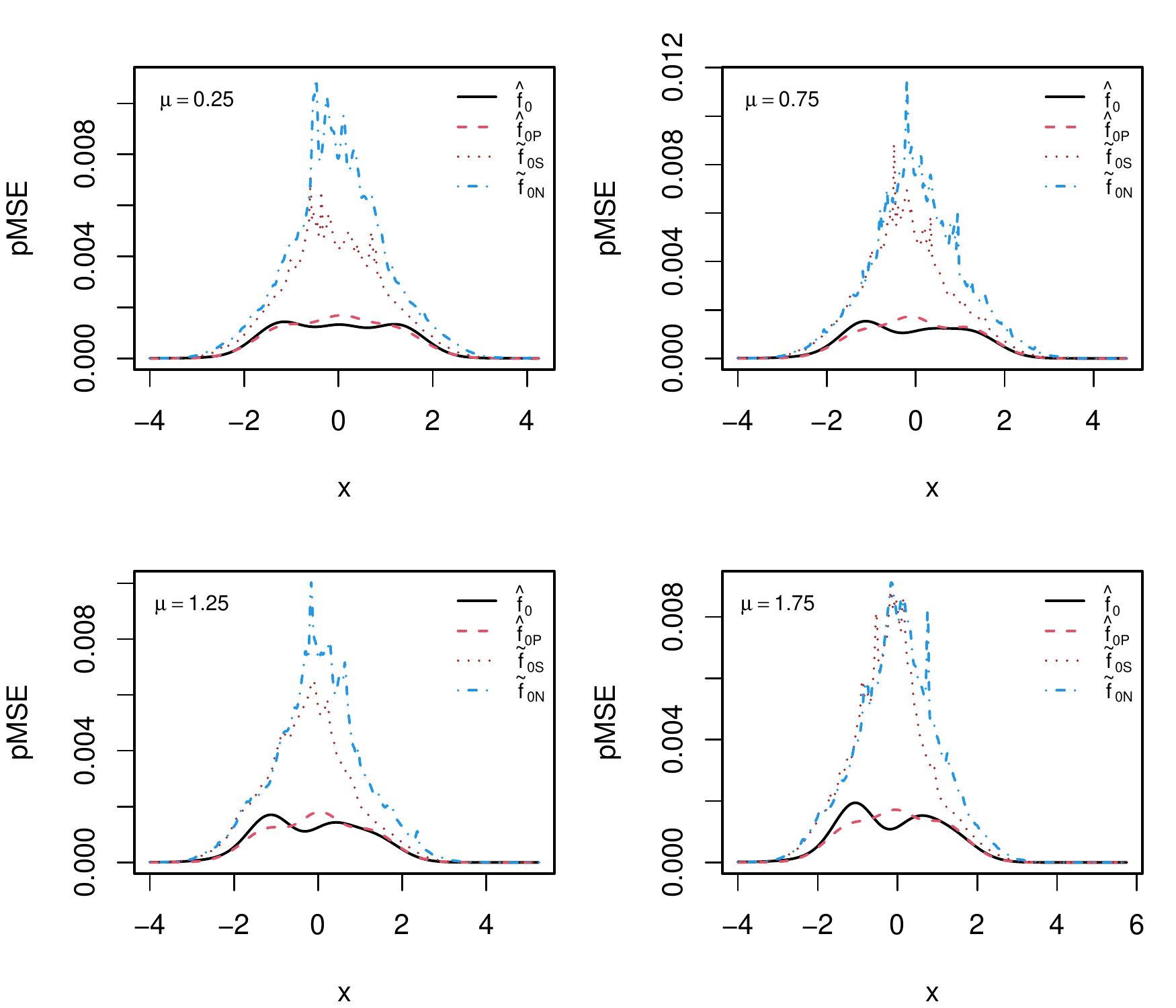}}
\caption{Simulated pointwise mean squared error of   the MABLE $\hat f_{0}$, the one-sample parametric MLE $\hat f_{0\mathrm{P}}$,
the semparametric kernel density estimate $\tilde f_{0\mathrm{S}}$,
and the one-sample nonparametric kernel density estimate $\tilde f_{0\mathrm{N}}$ of $f_0$ based 1000 datasets generated from normal distributions with $n_0=n_1=50$.}\label{fig: pmse norm}
\end{figure}
%
\begin{table}
\caption{Simulation results based on $B=1000$ Monte Carlo runs and samples of sizes $(n_0,n_1)$ from normal
distributions $\mathrm{N}(0, 1)$ and $\mathrm{N}(\mu,1)$ using optimal degree $\hat m$.}\label{tbl: simulation for normal distributions}
\begin{center}
\begin{tabular}{cccccccccccc}
  \hline\hline
 &  &  & \multicolumn{4}{c}{mse $\times 10^2$} &&
\multicolumn{4}{c}{mise $\times 10^4$}\\\cline{4-7}\cline{9-12}
$\mu$ & $E(\hat m)$ & $\sigma(\hat m)$ & $\hat\alpha_0$ &  $\hat\alpha_1$  & $\tilde\alpha_0$ & $\tilde\alpha_1$  &&
 $\hat f_{0\mathrm{P}}$   &  $\hat f_{0}$
&  $\tilde f_{0\mathrm{S}}$  &  $\tilde f_{0\mathrm{N}}$ \\
  &\multicolumn{10}{c}{$n_0=n_1=50$
  }&\\
0.25 & 15.00 & 2.77 & ~0.16 & ~4.22 & ~0.18 & ~4.61 && 6.46 & 6.24 & 16.22 & 24.93 \\
0.50 & 16.25 & 2.92 & ~0.54 & ~4.44 & ~0.65 & ~5.07 && 6.15 & 5.81 & 14.92 & 21.27 \\
0.75 & 17.29 & 2.91 & ~1.29 & ~5.06 & ~1.54 & ~6.06 && 5.94 & 5.57 & 15.86 & 21.31 \\
1.00 & 18.29 & 3.06 & ~2.52 & ~5.95 & ~3.07 & ~7.37 && 5.52 & 5.14 & 16.11 & 21.66 \\
1.25 & 19.31 & 3.28 & ~4.40 & ~7.24 & ~5.73 & ~9.49 && 5.68 & 5.65 & 14.87 & 19.65 \\
1.50 & 20.45 & 3.25 & ~6.99 & ~8.97 & 10.37 & 13.57 && 5.47 & 5.82 & 16.88 & 20.99 \\
1.75 & 21.47 & 3.40 & 10.46 & 10.37 & 15.60 & 16.19 && 5.48 & 5.69 & 17.19 & 19.66 \\
2.00 & 22.83 & 3.47 & 15.48 & 12.51 & 28.62 & 23.05 && 5.15 & 5.94 & 18.56 & 19.88 \\
  &\multicolumn{10}{c}{$n_0=n_1=100$
  }&\\
0.25 & 15.20 & 2.16 & 0.07 & 2.23 & ~0.08 & 2.32 && 3.05 & 3.52 & ~8.76 & 12.56 \\
0.50 & 16.28 & 2.49 & 0.28 & 2.33 & ~0.30 & 2.53 && 3.09 & 3.55 & ~8.18 & 11.87 \\
0.75 & 17.27 & 2.33 & 0.65 & 2.61 & ~0.72 & 2.77 && 2.84 & 3.26 & ~8.49 & 12.26 \\
1.00 & 18.42 & 2.58 & 1.20 & 3.03 & ~1.32 & 3.37 && 2.74 & 3.17 & ~9.30 & 12.26 \\
1.25 & 19.51 & 2.60 & 2.08 & 3.37 & ~2.34 & 3.84 && 2.78 & 3.11 & ~9.56 & 12.12 \\
1.50 & 20.60 & 2.75 & 3.69 & 4.49 & ~4.52 & 5.62 && 2.64 & 2.83 & ~9.22 & 11.56 \\
1.75 & 21.61 & 2.80 & 5.47 & 5.32 & ~6.73 & 6.60 && 2.54 & 2.93 & ~9.67 & 11.29 \\
2.00 & 23.02 & 2.94 & 8.17 & 6.02 & 11.15 & 8.32 && 2.45 & 3.01 & 10.09 & 11.61 \\
\hline\hline
\end{tabular}
\end{center}
\end{table}
%

{\bf Model 2: Exponential distributions} $X_0$ is exponential with  density $f_0(x)=e^{-x}$, $x>0$. $X_1$ is exponential with density
$f_1(x)=\mu^{-1}e^{-x/\mu}=f_0(x) e^{-\log\mu+(1-1/\mu)x}$, $x>0$, where $\mu = 1.25(0.25)3.00$ as in \cite{Qin-and-Zhang-2005}.
We choose $a=0$, $b=5\mu$, 
In this model, the bandwidths for $\tilde f_{i\mathrm{S}}$ and $\tilde f_{i\mathrm{N}}$ in Table \ref{tbl: simulation for exponential distributions} are those suggested by \cite{Qin-and-Zhang-2005}. The parametric MLE  is $\hat f_{i\mathrm{P}}(x)=f_0(x/\bar x_i)/\bar x_i$, where $\bar x_i$ is the sample mean of $x_{i1},\ldots,x_{in_i}$, $i=0,1$.
\begin{figure}
\centering
  \makebox{\includegraphics[width=5.8in]{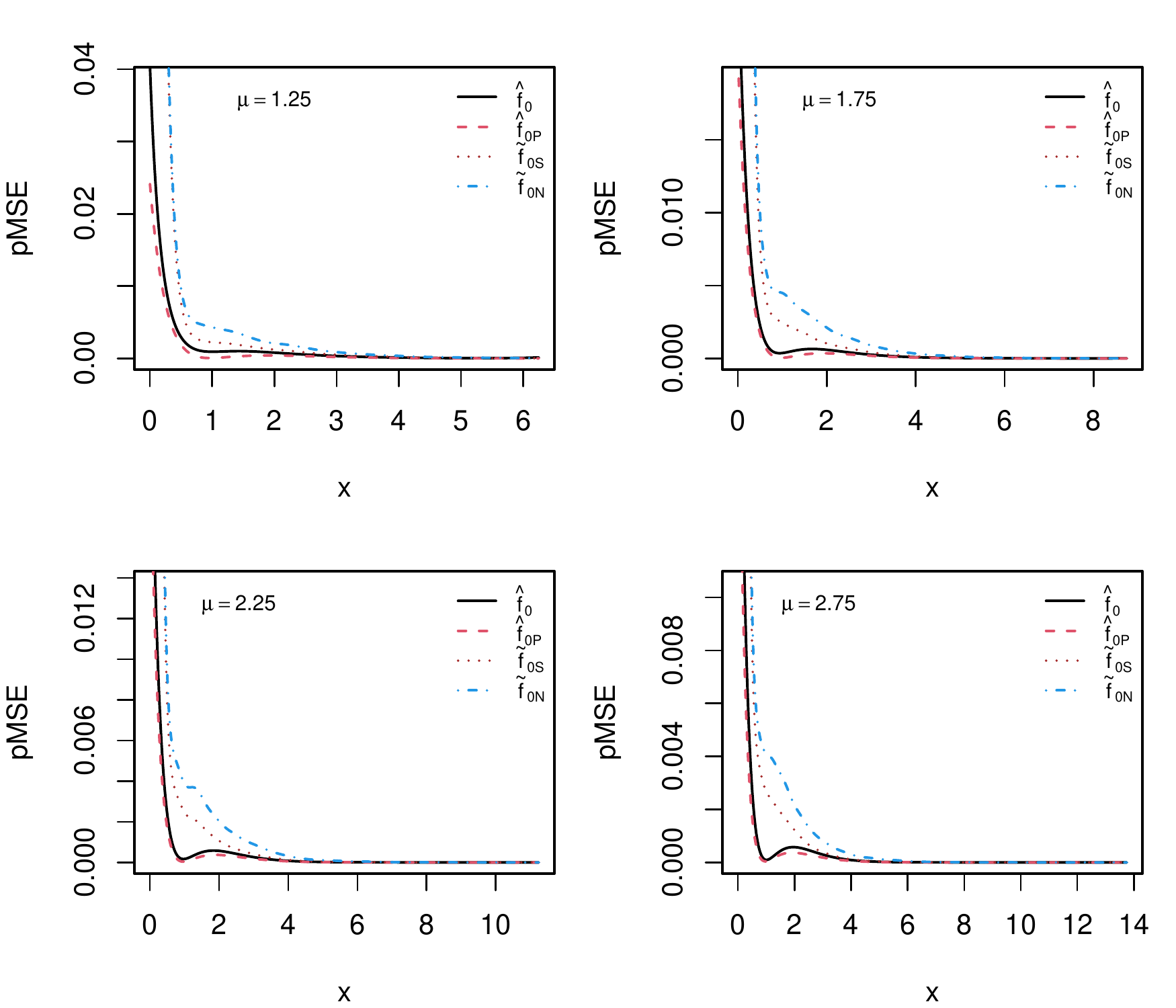}}
\caption{Simulated pointwise mean squared error of  the MABLE $\hat f_{0}$,  the one-sample parametric MLE $\hat f_{0\mathrm{P}}$,
the semparametric kernel density estimate $\tilde f_{0\mathrm{S}}$,
and the one-sample nonparametric kernel density estimate $\tilde f_{0\mathrm{N}}$ of $f_0$ based on 1000 datasets generated from exponential distributions with $n_0=n_1=50$.}\label{fig: pmse exp}
\end{figure}
\begin{table}
\caption{Simulation results based on $B=1000$ Monte Carlo runs and samples of sizes $(n_0,n_1)$ from exponential
distributions $\mathrm{Exp}(1)$ and $\mathrm{Exp}(\mu)$ using optimal degree $m=\tilde m$.}\label{tbl: simulation for exponential distributions}
\begin{center}
\begin{tabular}{cccccccccccc}
  \hline\hline
 &  &  & \multicolumn{4}{c}{mse $\times 10^2$} &&
\multicolumn{4}{c}{mise $\times 10^4$}\\\cline{4-7}\cline{9-12}
$\mu$ & $E(\tilde m)$ & $\sigma(\tilde m)$ & $\hat\alpha_0$ &  $\hat\alpha_1$  & $\tilde\alpha_0$ & $\tilde\alpha_1$  &&
 $\hat f_{0\mathrm{P}}$   &  $\hat f_{0}$
&  $\tilde f_{0\mathrm{S}}$  &  $\tilde f_{0\mathrm{N}}$ \\
  &\multicolumn{10}{c}{$n_0=n_1=50$
  }&\\
1.25 & 5.04 & 2.45 & 5.01 & 4.46 & 4.93 & 4.41 && 8.39 & 11.61 & 91.98 & 99.95 \\
1.50 & 4.87 & 1.96 & 5.07 & 4.01 & 5.29 & 4.10 && 7.00 & ~8.54 & 75.84 & 82.96 \\
1.75 & 4.72 & 1.81 & 4.94 & 3.42 & 5.81 & 3.79 && 5.59 & ~6.69 & 66.23 & 72.76 \\
2.00 & 4.68 & 2.25 & 5.03 & 3.53 & 6.33 & 4.06 && 5.82 & ~6.75 & 58.51 & 63.84 \\
2.25 & 4.70 & 2.54 & 4.64 & 3.06 & 6.46 & 3.75 && 4.82 & ~5.45 & 52.16 & 56.51 \\
2.50 & 4.66 & 2.81 & 4.71 & 3.20 & 7.46 & 4.16 && 4.42 & ~4.89 & 48.76 & 53.04 \\
2.75 & 4.54 & 2.20 & 4.54 & 3.13 & 7.28 & 3.99 && 4.21 & ~4.59 & 44.03 & 47.72 \\
3.00 & 4.58 & 2.87 & 5.25 & 2.96 & 8.99 & 4.14 && 3.39 & ~3.68 & 40.62 & 43.90 \\
  &\multicolumn{10}{c}{$n_0=n_1=100$
  }&\\
1.25 & 4.61 & 1.12 & 2.60 & 2.26 & 2.47 & 2.15 && 4.19 & 5.99 & 62.32 & 90.99 \\
1.50 & 4.46 & 1.05 & 2.32 & 1.87 & 2.48 & 1.88 && 3.62 & 4.55 & 58.10 & 80.45 \\
1.75 & 4.33 & 0.83 & 2.29 & 1.67 & 2.66 & 1.79 && 2.92 & 3.62 & 49.12 & 65.99 \\
2.00 & 4.22 & 0.75 & 2.36 & 1.66 & 3.00 & 1.86 && 2.73 & 3.09 & 44.31 & 58.99 \\
2.25 & 4.18 & 0.69 & 2.01 & 1.41 & 2.94 & 1.69 && 2.31 & 2.62 & 41.40 & 52.25 \\
2.50 & 4.15 & 0.37 & 2.40 & 1.64 & 3.50 & 1.95 && 2.34 & 2.44 & 38.76 & 48.49 \\
2.75 & 4.18 & 0.56 & 2.09 & 1.32 & 3.27 & 1.66 && 1.86 & 1.98 & 34.80 & 41.90 \\
3.00 & 4.22 & 0.76 & 2.45 & 1.53 & 4.01 & 2.03 && 1.97 & 2.05 & 33.19 & 40.12 \\
\hline\hline
\end{tabular}
\end{center}
\end{table}
The kernel density estimates suffers from serious boundary effect for a densities like expontial distribution. In the simulation presented in Figure \ref{fig: pmse exp} both $\tilde f_{i\mathrm{N}}$ and $\tilde f_{i\mathrm{S}}$ used the same bandwidth selected by the default method of R function ``density()''
which seems a little better than those selected by the method of \cite{Qin-and-Zhang-2005}.




From the above simulation results we observe the folloowing. (i) The optimal degree increases slowly as sample sizes increase; (ii) As sample sizes increase the variation of the optimal degree decreases; (iii) The larger $\alpha_1$ is the more eficient the proposed estimator $\hat\alpha_1$ is than $\tilde\alpha_1$; (iv) The proposed estimator $\hat f_0$ is very similar to the parametric one but is much better than the semiparamtric and the nonparametric ones.

\section{Large Sample Properties}\label{sect: large sample property}
We denote the chi-squared divergence($\chi^2$-distance) between densities $\varphi$ and $\psi$ by
$$\chi^2(\varphi\Vert\psi) =\int_{-\infty}^\infty\frac{[\varphi(y)-\psi(y)]^2}{\psi(y)} dy
\equiv \int_{-\infty}^\infty\Big[\frac{\varphi}{\psi}(y)-1\Big]^2 \psi(y)dy.
$$
We need the following assumptions for the asymptotic properties of  $\hat f_\B$ which will be proved in the appendix:
\begin{assump}\label{A1}
 There exists $\bm p_0\in \mathbb{S}_m$ 
and $k>0$ such that
$[{f_m(x;\bm p_0)-f_0(x)}]/{f_0(x)}=\mathcal{O}(m^{-k/2})$,
uniformly in $x\in[0,1]$, and thus $\chi^2(f_m(\cdot;\bm p_0)\Vert f_0)=\mathcal{O}(m^{-k})$.
\end{assump}
\begin{assump}\label{A2}
Assume that the zero vector $\bm 0\in\mathcal{A}$ and that the components of $\tilde r(x)$ are linearly independent.
\end{assump}

Let $C^{(r)}[0,1]$ be the class of functions which have $r$th continuous derivative $f^{(r)}$ on $[0,1]$.
If  $f_0\in C^{(r)}[0,1]$, and $f_0(x)\ge b_0>0$, $x\in [0,1]$, then Assumption \ref{A1} is fulfilled with $k=r$ \citep{Lorentz-1963-Math-Annalen}.

 A weaker sufficient condition can assure  Assumption \ref{A1}.
 A function $f$ is said to be {\em $\gamma$--H\"{o}lder
continuous} with $\gamma\in(0,1]$ if $|f(x)-f(y)|\le C|x-y|^\gamma$ for some constant $C>0$.
The following \citep[Lemma 3.1 of][]{Wang-and-Guan-2019} is a generalization of the result of \cite{Lorentz-1963-Math-Annalen} which requires a positive lower bound for $f_0$.
\begin{lemma}\label{thm: a sufficient existence condition for Bernstein model}
Suppose that $f_0(x)=x^a(1-x)^b\varphi_0(x)$ is a density on $[0,1]$, $a$ and $b$ are nonnegative real numbers,  $\varphi_0\in C^{(r)}[0,1]$, $r\ge 0$,  $\varphi_0(x)\ge b_0>0$, and $\varphi_0^{(r)}$ is $\gamma$-H\"{o}lder
continuous with $\gamma\in(0,1]$.
Then Assumption  \ref{A1} is fulfilled with $k=r+\gamma$.
\end{lemma}
We have the following asymptotic results in terms of distances $D^2_0(\bm\alpha,\bm p) =\chi^2(f_m(\cdot;\bm p)\Vert f_0)$ and $D^2_1(\bm\alpha,\bm p) =\chi^2(f_m(\cdot;\bm\alpha,\bm p)\Vert f_1)$.
\begin{theorem}\label{thm: large sample properties}
Under the density ratio model (\ref{eq: exponential tilting model}) and the assumptions  \ref{A1} with $k>0$,  and \ref{A2},
  as $n \to\infty $, with probability one  the  maximum
value of $\ell_m(\bm\alpha,\bm p)$ with $m={\cal O}(n^{1/k})$ is attained at $(\hat{\bm\alpha},\hat{\bm p})$ in the interior of
$\mathbb{B}_m(r_n)=\{(\bm\alpha,\bm p)\in \Theta_m(\mathcal{A})\,:\,D^2_i(\bm\alpha,\bm p) \le r_n,\, i=0,1\},$ 
where $r_n=n^{-1}\log n $. Thus the    mean $\chi^2$-distance between $f_m(\cdot;i\hat{\bm\alpha},\hat{\bm p})$ and $f_i(\cdot)$ satisfies
\begin{equation}\label{eq: convergence rate for contaminated data}
 \E[D^2_i(\hat{\bm\alpha},\hat{\bm p})]= \E\int\frac{[f_m(x;i\hat{\bm\alpha},\hat{\bm p}_m)-f_i(x)]^2}{f_i(x)}dx =\mathcal{O}\left(\frac{\log n}{n}\right),\; i=0,1.
\end{equation}
Moreover, almost surely, $\Vert \hat{\bm\alpha}-\bm\alpha_0\Vert^2 =\mathcal{O}(\log n/n)$.
\end{theorem}
\begin{rem}
Theorem \ref{thm: large sample properties} implies that 
$|\hat F_{i}(x)-F_i(x)|^2=\mathcal{O}(\log n/n)$,
uniformly on $[0,1]$, a.s., $i=0,1$.
\end{rem}


\section{Concluding Remark}\label{set: concluding remarks}
Unlike the empirical likelihood method of \cite{Qin-Zhang-Biometrika-2003,Qin-and-Zhang-2005} in which an estimate of a discrete probability mass function is obtained first then
smoothed  using  kernel method,  the proposed method produces smooth estimates of density and distribution functions directly. From the simulation study we also conclude that the proposed method
 does not only simply smooth the estimation but  also gives more accurate estimates. The improvement over the existing methods is significant especially for small samples.
 The proposed method also gives better estimates of coefficients of logistic regression for retrospective sampling data especially for samll sample data. Although the optimal model degree is large for some data the effective degrees of freedom, the number of nonzero mixing proportions $\hat p_i$, is usually much smaller.
Instead of the exponential tilting model (\ref{eq: exponential tilting model}), we can consider an even more general weighted model $f_1(x)=f_0(x)w(x; \bm\alpha),$
where  $w(x; \bm\alpha)$ is a known  nonnegative weight with unknown parameter $\bm\alpha$  and satisfies $\int w(x; \bm\alpha) f_0(x)dx=1$ and $w(x;\bm 0)=1$.

\section*{Appendix}
\label{sect: Appendix}

\subsection{Proof of Proposition \ref{prop: two-sample BPDR models are nested}}
\begin{proof}
For any $f_m(x;\bm\alpha,\bm p_m)\in\mathcal{D}_m(\mathcal{A})$, we have $f_m(x;\bm p_m)=\sum_{j=0}^m p_{mj}\beta_{mj}(x),$ and
$$\sum_{i=0}^mp_{mi}w_{mi}(\bm\alpha)=\int_0^1 f_{m}(x;\bm p_{m})\exp\{\bm\alpha^\T \tilde r(x)\}dx=1,$$
so that $f_m(x;\bm\alpha,\bm p_m)= f_m(x;\bm p_m)\exp\{\bm\alpha^\T \tilde r(x)\}$.
By Property 3.1. of \cite{Wang-and-Ghosh-2012-CSDA} or Lemma 2.2 of \cite{Guan-2017-jns} we also have that
$f_m(x;\bm p_m)=f_{m+1}(x;\bm p_{m+1})=\sum_{j=0}^{m+1} p_{m+1,j}\beta_{m+1,j}(x)$ with $p_{m+1,0}=(m+1)p_{m0}/(m+2)$, $p_{m+1,m+1}=(m+1)p_{mm}/(m+2)$, and
$p_{m+1,j}= \left[{j}p_{m,j-1}+{(m-j+1)}p_{mj}\right]/({m+2})$,  $j\in\ints(1,m).$
Thus we have
 $f_m(x;\bm\alpha,\bm p_m)
=f_{m+1}(x;\bm p_{m+1})\exp\{\bm\alpha^\T \tilde r(x)\} =f_{m+1}(x;\bm\alpha, \bm p_{m+1})$ and
$\sum_{i=0}^{m+1}p_{m+1,i}w_{m+1,i}(\bm\alpha)=1$. 
Hence $f_m(x;\bm\alpha,\bm p_m)\in \mathcal{D}_{m+1}(\mathcal{A})$.
\end{proof}
\subsection{Proof of Theorem \ref{thm: large sample properties}}
\begin{proof}
Let $\bm\alpha_0=(\alpha_{00},\ldots,\alpha_{0d})^\T$ be the true value of $\bm\alpha$ so that
 $
 \int_0^1 f_0(x) \exp\{\bm\alpha_0^\T\tilde r(x)\} dx=1.$
 By Assumption \ref{A1}, we have
\begin{equation}\label{eq: bernstein approx for fi}
f_m(x;i\bm\alpha_0, \bm p_0)=f_i(x)+R_m(x)\exp\{i\bm\alpha_0^\T\tilde r(x)\},\quad i=0,1,
\end{equation}
where $R_m(x)=f_0(x)\mathcal{O}(m^{-r/2})$. Thus
 $\int_0^1 f_m(x;\bm\alpha_0,\bm p_0)dx = 1 + \int_0^1 R_m(x)\exp\{i\bm\alpha_0^\T\tilde r(x)\} dx
 = 1+ \rho_m,$  where
$ \rho_m=\mathcal{O}(m^{-r/2})$. If we define $\tilde{\bm\alpha}_0=\tilde{\bm\alpha}_0{(m)}=(\tilde\alpha_{00},\alpha_{01},\ldots,\alpha_{0d})^\T$ with $\tilde\alpha_{00}
=\alpha_{00}-\log(1+\rho_m)$, then we have
$|\tilde{\bm\alpha}_0{(m)}-\bm\alpha_0|=|\log(1+\rho_m)|=\mathcal{O}(m^{-r/2})$,
$\int_0^1 f_m(x;\tilde{\bm\alpha}_0,\bm p_0)dx=\sum_{j=0}^m p_{0j}w_{mj}(\tilde{\bm\alpha}_0)= 1$, and
$$\frac{f_m(x;\tilde{\bm\alpha}_0,\bm p_0)-f_1(x)}{f_1(x)}
=\frac{R_m(x)}{(1+\rho_m)f_0(x)}-\frac{\rho_m}{1+\rho_m}=\mathcal{O}(m^{-r/2}).$$
Define the log-likelihood ratio
$\mathcal{R}(\bm\alpha,\bm p)=\ell(\bm\alpha_0,f_0) - \ell_m(\bm\alpha,\bm p)$.
Thus we have
\begin{eqnarray}\mathcal{R}(\bm\alpha,\bm p)
&=&-\sum_{i=0}^{1}\sum_{j=1}^{n_i}\log [f_m(x_{ij};i\bm\alpha,\bm p)/f_i(x_{ij})].
\end{eqnarray}
Consider subsets $$\Theta(\epsilon_0)=\{(\bm\alpha,\bm p)\in \Theta_m(\mathcal{A}):  \mbox{$\forall$  $x\in [0,1]$, $i=0,1$, $|f_m(x; i\bm\alpha,\bm p)/f_i(x)-1|\le \epsilon_0$}\},$$ $0<\epsilon_0<1$. Clearly, by \ref{A1} and (\ref{eq: bernstein approx for fi}),   $\Theta(\epsilon_0)$ is nonempty if $m$ is large enough.

By Taylor expansion we have, for $(\bm\alpha,\bm p)\in \Theta(\epsilon_0)$, $0<\epsilon_0<1$, and large $m$,
\begin{eqnarray*}
  \mathcal{R}(\bm\alpha,\bm p)&=& \sum_{i=0}^{1} \left\{\sum_{j=1}^{n_i} \Big[\frac{1}{2}U_{ij}^2(\bm\alpha, \bm p)-U_{ij}(\bm\alpha, \bm p)\Big] +\mathcal{O}(R_{mi}(\bm\alpha,\bm p))\right\},\, a.s.,
 \end{eqnarray*}
where    $U_{ij}(\bm\alpha,\bm p)=[{f_m(x_{ij};i\bm\alpha,\bm p)-f_i(x_{ij})}]/{f_i(x_{ij})}$, $j\in\ints(1,n_i)$, and $R_{mi}(\bm\alpha,\bm p)=\sum_{j=1}^{n_i}U_{ij}^2(\bm\alpha,\bm p)$, $i=0,1.$
Since $\E[U_{ij}(\bm\alpha,\bm p)]=0$, $\sigma^2[U_{ij}(\bm\alpha,\bm p)]=\E[U_{ij}^2(\bm\alpha,\bm p)]=D^2_i(\bm\alpha,\bm p)$,
by the LIL we have
$\sum_{j=1}^n U_{ij}(\bm\alpha,\bm p)/\sigma[U_{ij}(\bm\alpha,\bm p)]=\mathcal{O}(\sqrt{n\log\log n})$, a.s..
By the strong law of large numbers we have, a.s.,
\begin{equation}\label{log likelihood ratio fm(p)}
\mathcal{R}(\bm\alpha,\bm p) =\sum_{i=0}^1\left\{\frac{n}{2} D^2_i(\bm\alpha,\bm p) - \mathcal{O}(D_i(\bm\alpha,\bm p)\sqrt{n\log\log n}) +o(nD^2_i(\bm\alpha,\bm p))\right\}.
\end{equation}
If $D^2_i(\bm\alpha,\bm p)= r_n=\log n/n$, then,  by (\ref{log likelihood ratio fm(p)}),
there is an $\eta>0$ such that
 $\mathcal{R}(\bm\alpha,\bm p)\ge
\eta \log n,\, a.s.. $
If $(\bm\alpha,\bm p)=(\tilde{\bm\alpha}_0,\bm p_0)$ and $m=Cn^{1/k}$, we have $D^2_i(\tilde{\bm\alpha}_0,\bm p_0)=\mathcal{O}(m^{-k})=\mathcal{O}(n^{-1})$.
By (\ref{log likelihood ratio fm(p)}) again we have
 $\mathcal{R}(\tilde{\bm\alpha}_0,\bm p_0)=
\mathcal{O}(\sqrt{\log\log n})$, a.s..
Therefore, similar to the proof of Lemma 1 of \cite{QinLawless}, we have
\begin{equation}\label{eq: convergence rate of Di-sqr}
D^2_i(\hat{\bm\alpha},\hat{\bm p})
=\int_{0}^1\frac{[f_m (x;i\hat{\bm\alpha},\hat{\bm p})-f_i(x)]^2}{f_i(x)}dx\le\frac{ \log n}{n}, a.s. ,
\end{equation}
and $(\hat{\bm\alpha},\hat{\bm p})\in\Theta(\epsilon_0)$.
Thus (\ref{eq: convergence rate for contaminated data}) follows.
Define
$$  \psi({\bm\alpha}, g) =    \int_{0}^1\frac{g^2 (x;{\bm p})}{f_0^2(x)}\left[\frac{w(x;{\bm \alpha})}{w(x;{\bm \alpha}_0)}-1\right]^2f_1(x)dx, $$
where $w(x;{\bm \alpha})=\exp\{\bm\alpha^\T\tilde r(x)\}$ and $g$ is a density on $[0,1]$.
Then we have
\begin{eqnarray*}
  \psi(\hat{\bm\alpha},\hat{f}_0)
    &\le & 2D^2_1(\hat{\bm\alpha},\hat{\bm p})+2\int_{0}^1\left[\frac{f_m (x;\hat{\bm p})}{f_0(x)}-1\right]^2f_1(x)dx\\
    &\le & 2D^2_1(\hat{\bm\alpha},\hat{\bm p})+2CD^2_0(\hat{\bm\alpha},\hat{\bm p})\\
    &=&\mathcal{O}(\log n/n),
\end{eqnarray*}
where $C=\max_{x\in [0,1]} w(x;{\bm \alpha}_0)$. It is clear that $\psi
(\bm\alpha_0, g)=0$, $\dot \psi
(\bm\alpha_0, g)=0$ and
 $$\ddot \psi
(\bm\alpha_0, f_0)=2\int_{0}^1 \tilde r(x)\tilde r^\T(x) f_1(x)dx \equiv 2J(\bm\alpha_0).$$
By Taylor expansion  and (\ref{eq: convergence rate of Di-sqr}) we have
$\psi(\hat{\bm\alpha},\hat{f}_m)= (\hat{\bm\alpha}-\bm\alpha_0)^\T J(\bm\alpha_0)(\hat{\bm\alpha}-\bm\alpha_0)+o(R_{n}),
$
where
$R_{n}=\Vert \hat{\bm\alpha} -\bm\alpha_0\Vert^2 +\mathcal{O}(\log n/n).$
Then we have
$(\hat{\bm\alpha}-\bm\alpha_0)^\T J(\hat{\bm\alpha}-\bm\alpha_0)+o(\Vert \hat{\bm\alpha} -\bm\alpha_0\Vert^2)=\psi(\hat{\bm\alpha},\hat{f}_m)+o(\log n/n)$
and thus $ (\lambda_0+o(1))\Vert \hat{\bm\alpha} -\bm\alpha_0\Vert^2 \le \mathcal{O}(\log n/n)$, where $\lambda_0$ is the minimum eigenvalue of $J(\bm\alpha_0)$.
Because the components of $\tilde r(x)=(1, r^\T(x))^\T$ are linearly independent,
$J(\bm\alpha_0)$ is positive definite.
 Thus  $\lambda_0>0$ and we have
$\Vert \hat{\bm\alpha} -\bm\alpha_0\Vert^2 = \mathcal{O}(\log n/n)$, a.s..
 The proof is complete.
 \end{proof}

\bibliographystyle{Chicago}


\def\polhk#1{\setbox0=\hbox{#1}{\ooalign{\hidewidth
  \lower1.5ex\hbox{`}\hidewidth\crcr\unhbox0}}} \def\cprime{$'$}

\end{document}